\documentclass[journal]{vgtc}                

\ifpdf
  \pdfoutput=1\relax                   
  \usepackage{graphicx}                
  \DeclareGraphicsExtensions{.pdf,.png,.jpg,.jpeg} 
\else
  \usepackage{graphicx}                
  \DeclareGraphicsExtensions{.eps}     
\fi%

\graphicspath{{figures/}{pictures/}{images/}{./}} 

\usepackage{microtype}                 
\PassOptionsToPackage{warn}{textcomp}  
\usepackage{textcomp}                  
\usepackage{mathptmx}                  
\usepackage{times}                     
\usepackage{cite}                      
\usepackage{tabu}                      
\usepackage{booktabs}                  
\usepackage{amsmath}
\usepackage{subfigure}

\usepackage{amssymb}
\usepackage{amsmath}
\usepackage{graphicx}
\usepackage[table,xcdraw]{xcolor}
\usepackage{booktabs}
\usepackage{multirow}
\usepackage[super]{nth}

\usepackage{lscape}

\onlineid{0}

\vgtccategory{Research}
\vgtcpapertype{please specify}

\title{A Comparative Analysis of Virtual Reality Head-Mounted Display Systems}

\author{Arian Mehrfard, Javad Fotouhi, Giacomo Taylor, Tess Forster, Nassir Navab, and Bernhard Fuerst}
\authorfooter{
\item
 Arian Mehrfard, Javad Fotouhi, and Nassir Navab are with the Laboratory for Computational Sensing and Robotics, at The Johns Hopkins University, Baltimore, Maryland, United States. E-mail: \{mehrfard, javad.fotouhi, nassir.navab\}@jhu.edu
\item
 Giacomo Taylor, Tess Forster, and Bernhard Fuerst are with Verb Surgical Inc., Mountain View, California, United States. E-mail: \{giacomotaylor, tessforster, benfuerst\}@verbsurgical.com
}

\shortauthortitle{Biv \MakeLowercase{\textit{et al.}}: Global Illumination for Fun and Profit}

\abstract{
With recent advances of Virtual Reality (VR) technology, the deployment of such will dramatically increase in non-entertainment environments, such as professional education and training, manufacturing, service, or low frequency/high risk scenarios. Clinical education is an area that especially stands to benefit from VR technology due to the complexity, high cost, and difficult logistics. The effectiveness of the deployment of VR systems, is subject to factors that may not be necessarily considered for devices targeting the entertainment market. In this work, we systematically compare a wide range of VR Head-Mounted Displays (HMDs) technologies and designs by defining a new set of metrics that are \textit{1)} relevant to most generic VR solutions and \textit{2)} are of paramount importance for VR-based education and training. We evaluated ten HMDs based on various criteria, including \textit{neck strain}, \textit{heat development}, and \textit{color accuracy}. Other metrics such as \textit{text readability}, \textit{comfort}, and \textit{contrast perception} were evaluated in a multi-user study on three selected HMDs, namely \textit{Oculus Rift S}, \textit{HTC Vive Pro} and \textit{Samsung Odyssey+}. Results indicate that the \textit{HTC Vive Pro} performs best with regards to comfort, display quality and compatibility with glasses.}

\keywords{Virtual reality, Head-mounted display, Enterprise VR use}

\CCScatlist{ 
 \CCScat{K.6.1}{Management of Computing and Information Systems}%
{Project and People Management}{Life Cycle};
 \CCScat{K.7.m}{The Computing Profession}{Miscellaneous}{Ethics}
}

\teaser{
  \centering
  \includegraphics[width=\linewidth]{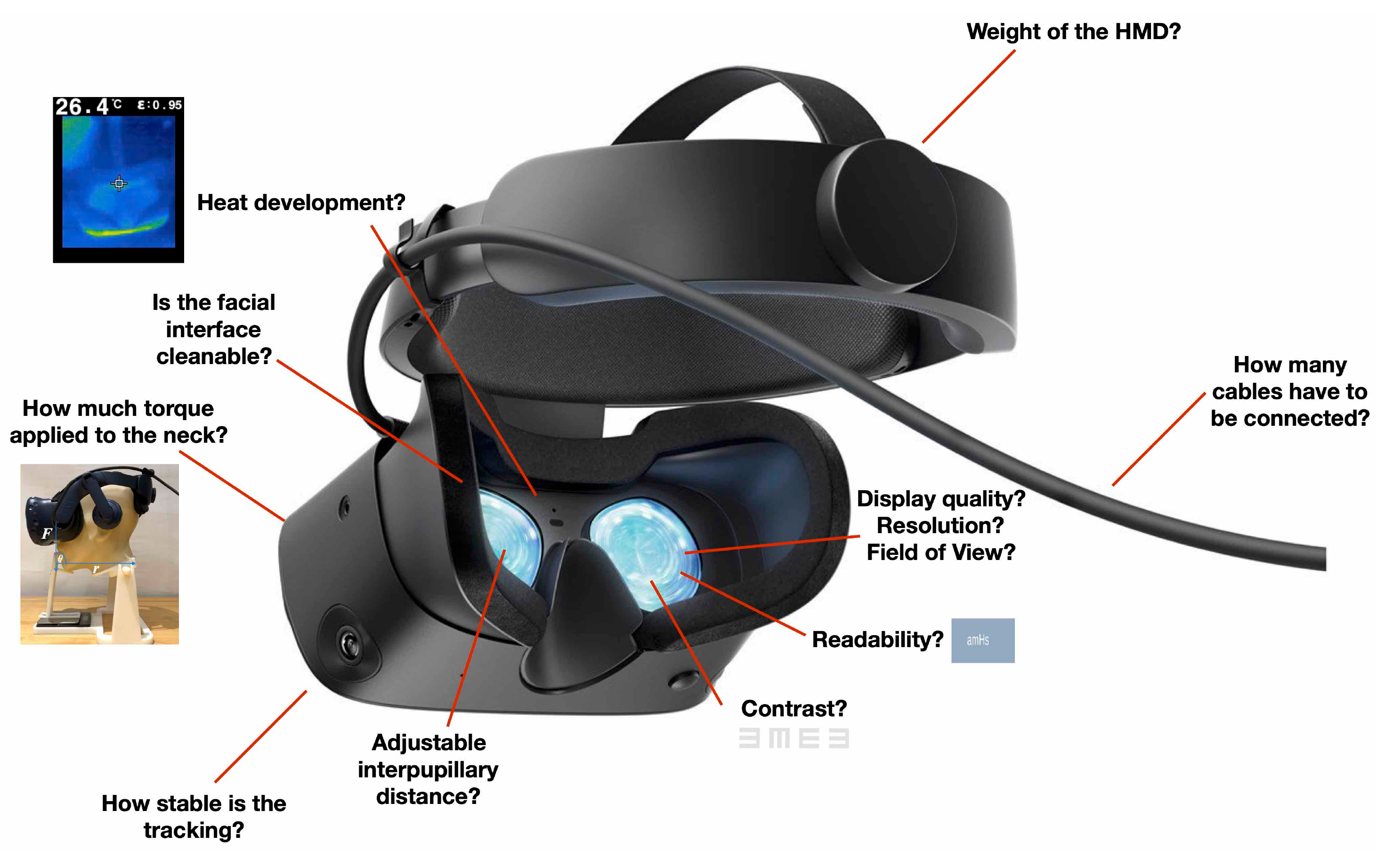}
  \caption{Overview of the VR metrics}
	\label{fig:teaser}
}

\renewcommand{\manuscriptnotetxt}{}

\vgtcinsertpkg

\begin{document}
\firstsection{Introduction}
\maketitle
Benefits of virtual reality (VR) systems were recognized as early as the 1960s, and systems have been deployed in professional settings since then. However, these early systems were very task-specific and costly, preventing wide-scale adoption.
In \cite{sutherland1968head}, Sutherland discusses his early tracked 3D head-mounted display (HMD) and reports the favorable responses of users. Sutherland's early prototype was the first device that resembles current, immersive VR headsets, and many consider his work the foundation of modern VR and augmented reality.
Since then, there have been many scientific works investigating the properties and the feasibility of VR for different applications. Psotka (1995) discusses the benefits of VR and many of the cognitive factors that affect immersion in virtual environments in \cite{psotka1995immersive}. In \cite{youngblut1998educational}, Youngblut (1998) reports on past VR applications involving general education of students.

Mazuryk and Gervautz (1999) give a detailed perspective on VR, defining its benefits and challenges, and outline possible applications the technology allows for. At its core, VR constitutes a human computer interface that allows users to interact with a virtual environment through natural, real world motions. Utilizing this interface allows for finer control required in modeling and design applications, as well as intuitive collaboration in interactive, virtual environments \cite{Mazuryk_virtualreality, mujber2004virtual}. VR can also allow for quick prototyping of designs or visualization of large amounts of data such as architectural walk-throughs or design of large machines such as airplanes. In \cite{mujber2004virtual}, Mujber states that the introduction of VR into manufacturing processes leads to an improved development pipeline. Definition, modeling, and verification can be done in an virtual environment which generates information about the structure and behaviour of objects, improving the product design and saving costs.

The first instance of VR training is described in \cite{baumann1993military}, where VR was developed for military simulations and telepresence training. Since then VR has been deployed in astronautics, flight training, medicine and other civilian fields \cite{khor2016augmented, hoffman1997virtual}.

In \cite{yavrucuk2011low} Yavrucuk describes a helicopter simulator using a VR HMD, and cited the benefits of a 360\textdegree view in the cockpit, intuitive interaction with the control and the substantially lower cost compared to conventional flight simulators.

VR technologies have also been applied to general education in tertiary, and in rare cases also secondary education, as reported by Freina and Ott in \cite{freina2015literature}. They state that VR was used in university and pre-university teaching for scientific subjects where a virtual world is used to visualize complex physical and chemical concepts. In \cite{hoffman1997virtual} Hoffman suggests using the same concepts for medical training at the university level. Another large benefit of VR is its ability to be used in professional training that can be logistically difficult or hazardous to conduct in the real world. Two examples of this are corrosion prevention and control training for the U.S. Army \cite{webster2014corrosion}; and rehabilitation of people with intellectual disabilities where the safe and accessible environment of VR offers desirable benefits to the real world \cite{standen2005virtual}. Another example can be found in \cite{sacks2013construction}, where VR is utilized to improve construction safety training by simulating hazardous conditions. The results of this study suggest that the VR based training results in an increase in workers' attention and increased engagement, ultimately leading to better training outcomes.

Furthermore, VR's natural interface can be utilized to teleoperate robots, allowing for human dexterity and decision making in hazardous remote environments without operator endangerment. Similarly, telepresence can also be used for observation or mentoring in cases where the remote location might be too hazardous, far away or unreachable for a human \cite{khor2016augmented}.

Many examples of VR applications for training come from medicine. Surgical training in particular represents an area that is high risk and cost ineffective, where patient safety and difficult other logistics pose significant obstacles. In \cite{alaker2016virtual}, Alaker conducts a systematic review on VR training in laparoscopic surgery evaluating VR training against conventional trainers and finds that VR training can be significantly more effective than video training and comparable to more sophisticated box trainers. In \cite{gallagher2005virtual}, Gallagher discusses the employment of VR in surgical skills training stating that VR improves the state of training for minimally invasive surgery but that it has to be thoughtfully implemented in the education program in order to be successful.


VR training is particularly effective for high risk or hazardous environments, or in  scenarios where accurate reproduction of complex stimuli is difficult. In these cases, VR-based training and visualization is often superior to other training methods, offering improved immersion, presence, and spatial awareness. In teleoperation tasks, these benefits have been shown to facilitate more intuitive and dexterous control, leading to safer and more efficient outcomes. For training, VR simulation can lead to higher engagement of students and knowledge retention \cite{freina2015literature, sacks2013construction}.

\begin{figure}[tb]
 \centering 
 \includegraphics[width=\columnwidth]{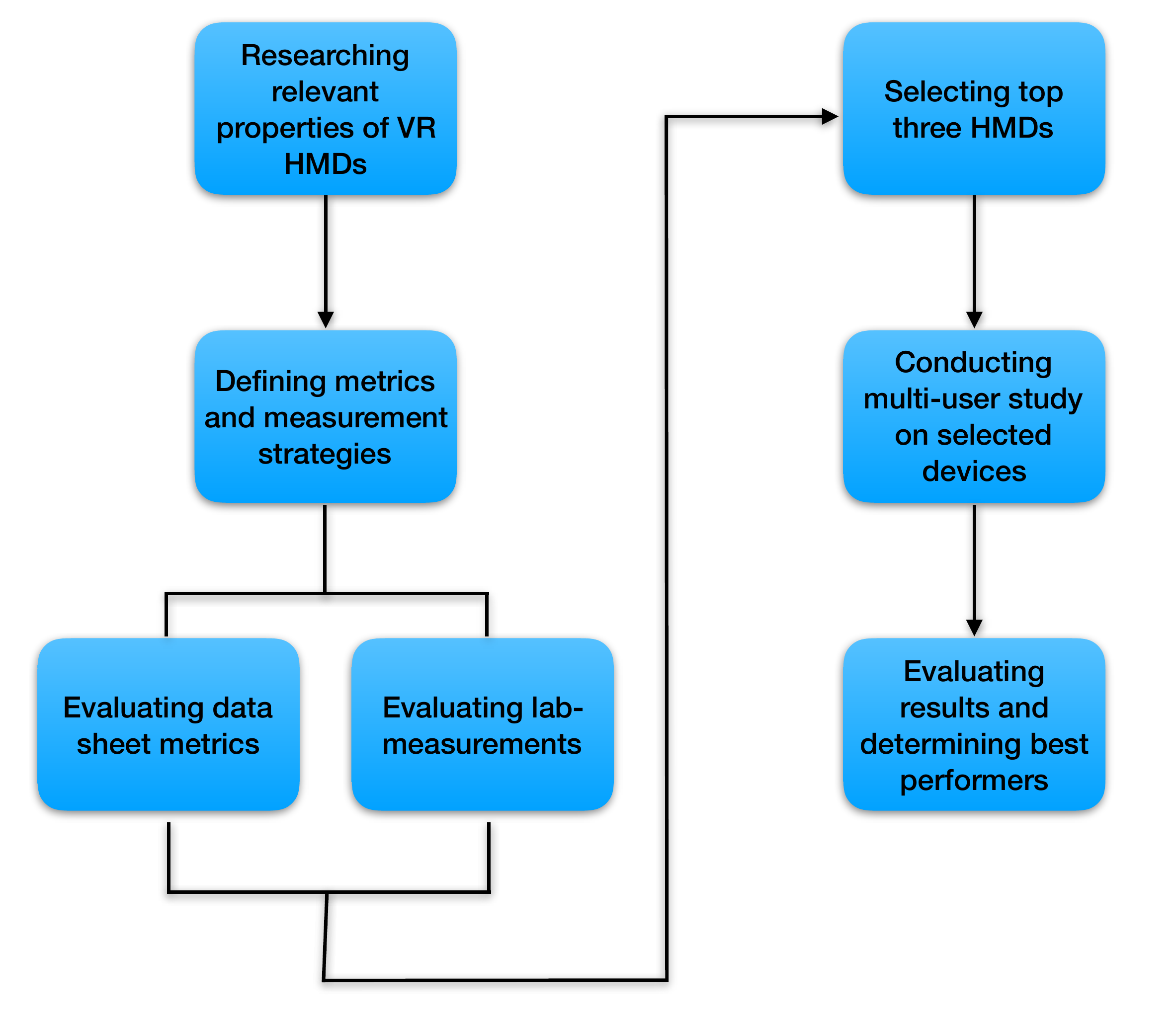}
 \caption{Strategy to define metrics and evaluete HMDs}
 \label{fig:Flowchart}
\end{figure}

According to \cite{Mazuryk_virtualreality, mujber2004virtual}, much of the VR technologies discussed above have historically been prohibitively expensive, ranging from \$800 for low quality devices to \$1 million for military HMDs. It is only recently that advances in VR HMD technology that high quality and usable HMDs are commercially and economically accessible by the general public, and therefore available for research, professional and large scale enterprise deployment \cite{slater2016enhancing, khor2016augmented, freina2015literature}. In 2016, Anthes et al. overview the state of VR technology, comparing various types of VR HMDs and their limitations. The authors describe recent advances in VR technology as "the second wave of VR", and most notably compare wired HMDs and mobile HMDs \cite{anthes2016state}.

Due to the limited computing power and diversity in current mobile VR HMDs, we are only aware of various HMDs that use an inserted phone as a display and the \textit{Oculus Quest}, we only consider wired HMDs in our work.

Most commercially available VR devices today are predominantly designed for the entertainment market. VR HMDs designed for gaming and entertainment ultimately prioritize a set of factors different from those for a professional market. For commercial applications then, it is increasingly challenging to select the appropriate VR HMD from the number of commercially available VR systems. 
In this work we define a new set of metrics relevant to most VR applications, but specifically crucial for VR-based education and training.
We then choose and systematically evaluate three commercially available VR HMDs according to these metrics, our approach is shown in Fig.~\ref{fig:Flowchart}.
\section{Method}\label{sec:method}
We define the evaluation metrics based on the following fundamental properties required for VR solutions:

\paragraph{\textbf{Image quality}} correlates with the level of details that the user perceives and impacts immersion and presence which are the core concepts of VR that give users the impression of \textbf{being} in a virtual environment. In a skill training VR solution, the quality of the visual experience and level of immersion widely associate with the transfer of skills to the real world and are therefore of utmost importance\cite{freina2015literature, sacks2013construction, bowman2007immersion}.

\paragraph{\textbf{User comfort}} is influenced by the \textit{weight} and balance of HMD on the user's head, heat development, and tracking stability among others. Discomfort can result in user dissatisfaction, hence reduce the skill transfer and overall willingness to use VR.

\paragraph{\textbf{Secondary features}} encompass many features of the VR design which do not necessarily fall into the the above categories, but influence the usability of the HMD. The main considerations here are the VR HMD \textit{integrated audio}, \textit{connections ports}, \textit{setup complexity}, \textit{setup time}, and its \textit{hygiene}.

Depending on the measuring approach, each metric is assigned to one of the following three categories. The first category includes the information taken from the \textit{data-sheet} and facts provided by the manufacturer that establish a comparative baseline. The second category are \textit{measurements}, properties of the HMD that were measured through various instruments and devices. The third category is the \textit{user study}, which is evaluated through a multi-user study and is quantified based on users' performance and feedback. 

We chose ten VR HMD technologies which encompass most commercially available VR HMDs in May 2019, and evaluated them against the metrics in categories one and two. Following this evaluation, the three best performing devices were selected for in-depth evaluation through a multi-user study.

\subsection{Data sheet}
\paragraph{\textit{Field of view (image quality):}} Field of View (FoV) describes the extent of the virtual environment that is visible through the HMD, \textit{i.e.} the angle of view from the users eye to the lens. Higher\textit{FoV}is associated with higher immersion as the user is able to perceive more of the virtual world~\cite{lin2002effects, bowman2007immersion}. This measure is not constant for HMDs that allow users to change the distance between the eye and lens.
    
\paragraph{\textit{Resolution (image quality)}:} \textit{Resolution} is a measure of the amount of pixels in the display. Higher \textit{resolution} correlates directly with the quality of the visual perception, immersion, and the level of details~\cite{bowman2007immersion, ziefle1998effects}. Higher display resolutions also associate with improved text readability in virtual environments. 

\paragraph{\textit{Interpupillary distance adjustablity (image quality and comfort)}:} The quality of the stereo vision is contingent on the correct alignment of the lenses of the HMD with the pupils of the user. In humans, the average adult interpupillary distance (IPD) is $63$\,mm, with the majority of adults having IPDs between $50$\,mm and $75$\,mm~\cite{dodgson2004variation}. 
Misalignment leads to decreased quality stereo vision, diffuses the rendered image, and can result in cyber sickness or headache. Considering the deployment of the HMDs across a broad range of the population, it is critical that modern HMD systems become equipped with variable IPD\cite{kooi2004visual, mon1993binocular}.
    
\paragraph{Integrated audio (\textit{secondary features}):} An \textit{integrated audio} system eliminates the effort for mounting a peripheral audio device such as headphones. External headphones require additional cables and can interfere with the ergonomic head-strap, if they do not optimally fit on the HMD, user comfort can be significantly decreased. \textit{Integrated audio} technologies are further sub-categorized to \textit{1)} earpieces that block substantial amount of background sound, and \textit{2)} open sound systems that do not block any real-world sounds. For the evaluation in this manuscript, we only considered the sole availability of the audio system, as there is not enough research present to characterize the benefits and drawbacks of different audio technologies in VR devices. 
    
\paragraph{Connectivity (\textit{secondary features}):} The effort of setting up a VR system is associated with the amount of cables and the special ports that are required.

\subsection{Measurements}

\paragraph{Brightness and color accuracy (\textit{image quality}):} A brighter screen is associated with a higher quality display and viewing experience. The measured unit for \textit{brightness} is nits, where nit is expressed as candle per square meter $\frac{cd}{m^2}$. The \textit{color accuracy} is measured as the distance between the intended and the true displayed color $\Delta\,e$ in the \textbf{CIELAB color space}. These two metrics were measured with \textit{Datacolor SpyderX Elite (Datacolor, Luzern)} display calibration device. To perform the measurements on the VR systems, we mirrored the computer desktop to the HMD using \textit{SteamVR (Valve Corporation, Bellevue, WA, US)} software, enabling us to employ the calibration on the HMD displays. For each measurement, the calibration device was fixated onto one of the lenses of the HMD at an orientation such that the virtual camera is continuously pointed at the calibration area.

\paragraph{Heat management (\textit{comfort}):} Excessive heat development on the device can lead to severe discomfort. We measure the surface heat of the VR headset with \textit{FLIR TG165 (FLIR, Wilsonville)} thermal camera, and we target the hottest areas around the facial interface. In order to measure consistently among all devices, each HMD was first turned off and disconnected for at least $30$ minutes. Next, we mounted each HMD on a mannequin's head and let a VR environment run, confirming that the proximity sensor is engaged and the displays are turned on. Consequently, we measured the hottest areas of the HMD using the thermal camera.
    
\paragraph{Total weight (\textit{comfort}):} Heavier HMDs exert more stress on the head of the wearer, factoring into higher discomfort and exhaustion. The weights of all HMDs were measured and compared using a digital scale.

\begin{figure}[tb]
 \centering 
 \includegraphics[width=\columnwidth]{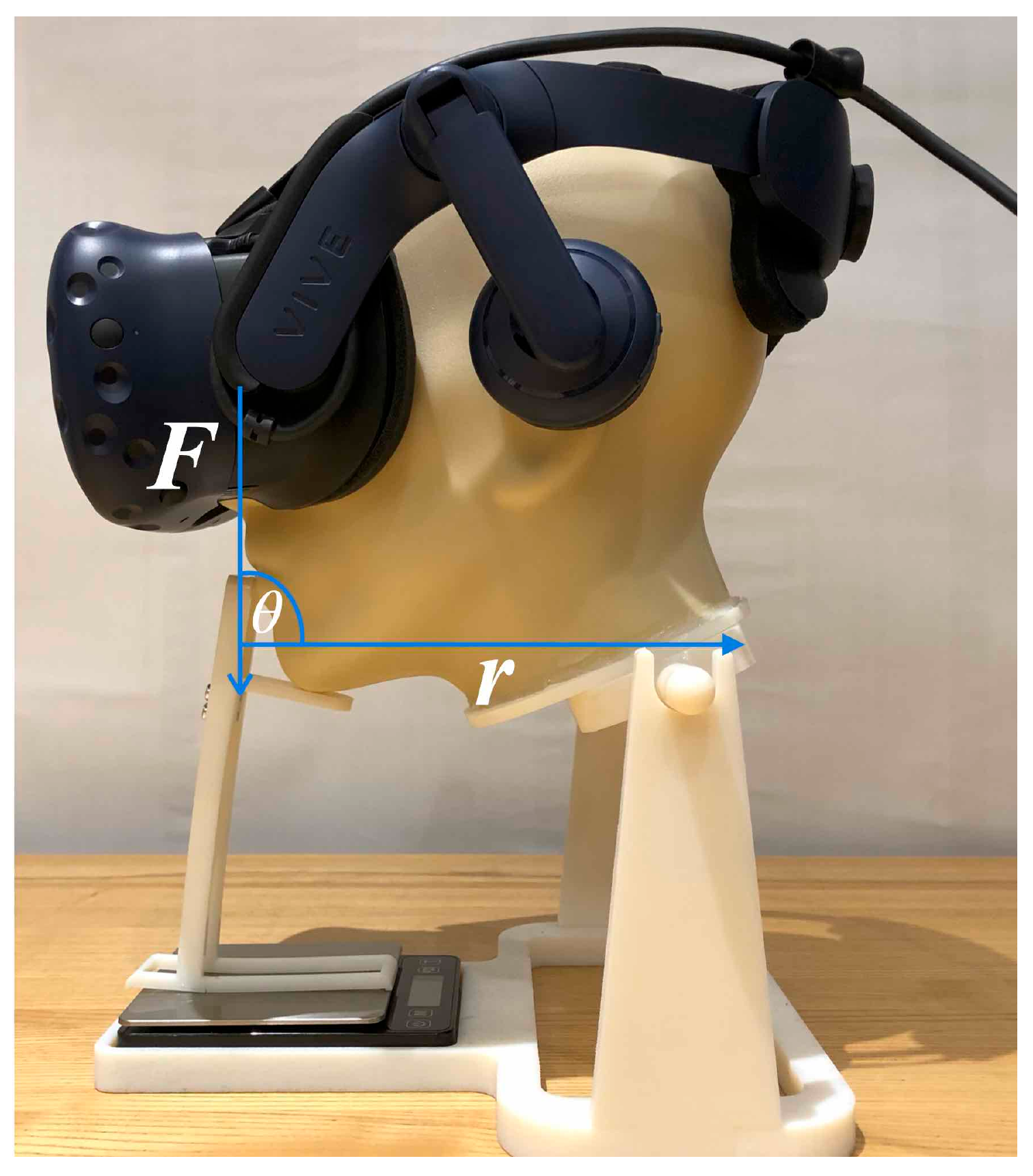}
 \caption{Setup to measure the torque applied to the neck}
 \label{fig:torque}
\end{figure}

\paragraph{Neck strain (\textit{comfort}):} Corresponding to the HMD's total weight, the device applies extra force to the front of the face, thereby creates \textit{torque} around the neck. Fig.\ref{fig:torque} illustrates the setup that we designed to approximate the \textit{torque} around the neck.

\begin{equation}
     \mathbf{\tau} = \mathbf{F \times r},
\end{equation}
where $\mathbf{F}$ is the force vector pushing the HMD down, and $\mathbf{r}$ is the distance vector between the chin and the axis of rotation. We approximated the angle between the two vectors to be $90$\textdegree ~in our setup, hence resulting in a simplified formula $\mathbf{\tau} = |\mathbf{F}| \, |\mathbf{r}|$.

\paragraph{Hygiene \textit{(secondary features)}:} Cleanability is an important property to consider for any HMD that is to be deployed in a large scale. Due to the direct contact between the facial interface and the users' skin, it is of paramount importance in applications that the device is shared among various users that each person receives the device in a clean and sanitary state. The device can either be disinfected, or the facial interface replaced. Therefore, we consider two factors in our analysis, \textit{1)} if the material repels liquids thus being \textit{wipeable}, and \textit{2)} the replaceability of the facial interface or availability of third party covers.
    
\subsection{Multi-user study}
\begin{figure}[tb]
 \centering 
 \includegraphics[width=\columnwidth]{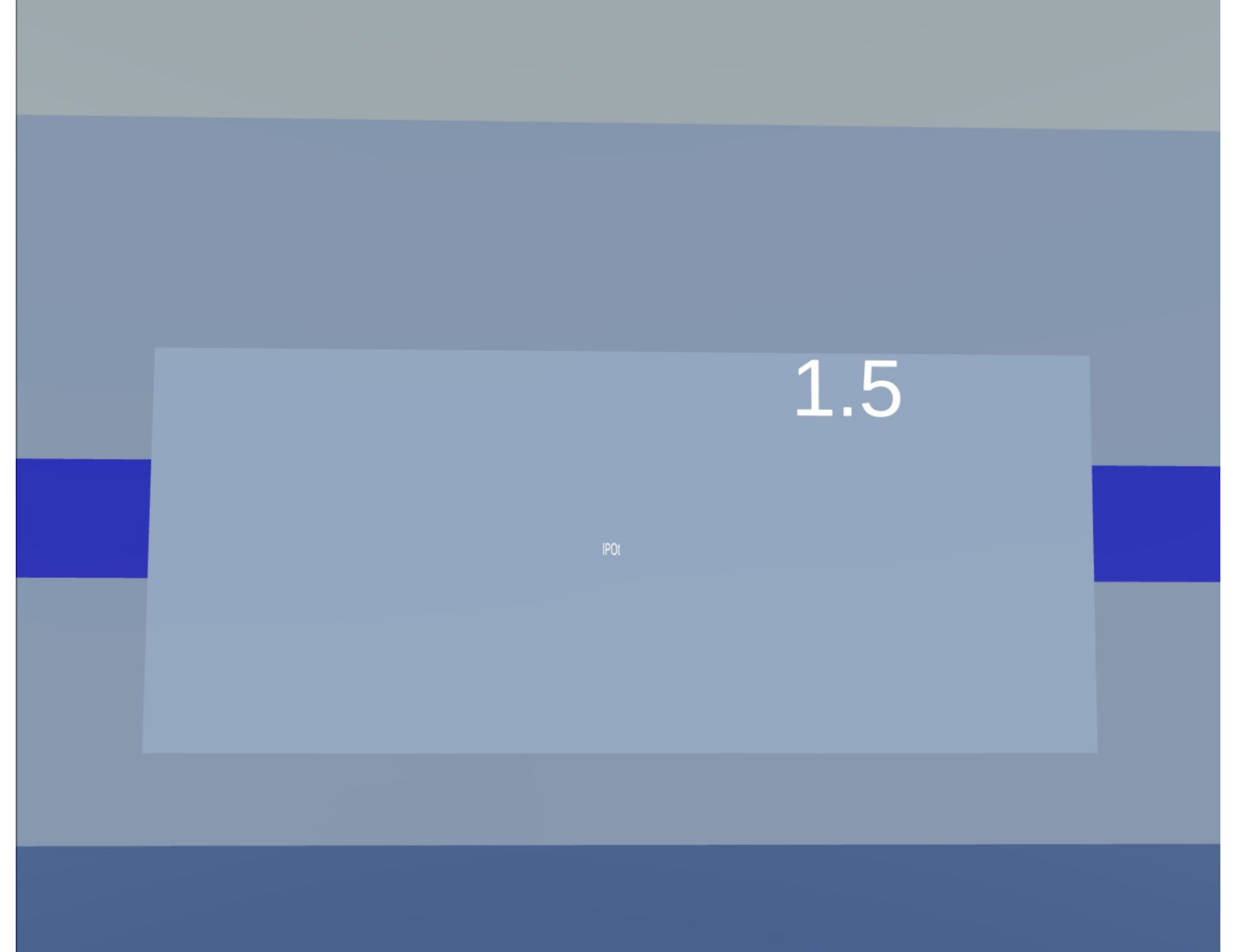}
 \caption{The readability test, with the text in the center and font size displayed at the top right}
 \label{fig:ReadScreenshot}
\end{figure}

\paragraph{Text readability \textit{(image quality)}:} Written instructions are one of the major media of communication to users in VR, therefore it is important that virtual text content appear well readable. To characterize this property, we designed a virtual environment (VE) in which the user stands a fixed distance away from a given text, and is asked to read the text content out loud. The text consists of four random characters containing upper and lower case letters and digits (Fig. \ref{fig:ReadScreenshot}). The users are tasked with reading out the characters and reducing the font size, in decrements of $0.05$, until they are unable to read them or misread them. Due to the random component, each user repeats the test three times. In this setting, a lower font size indicates better text readability in the HMD.

\begin{figure}[tb]
 \centering 
 \includegraphics[width=\columnwidth]{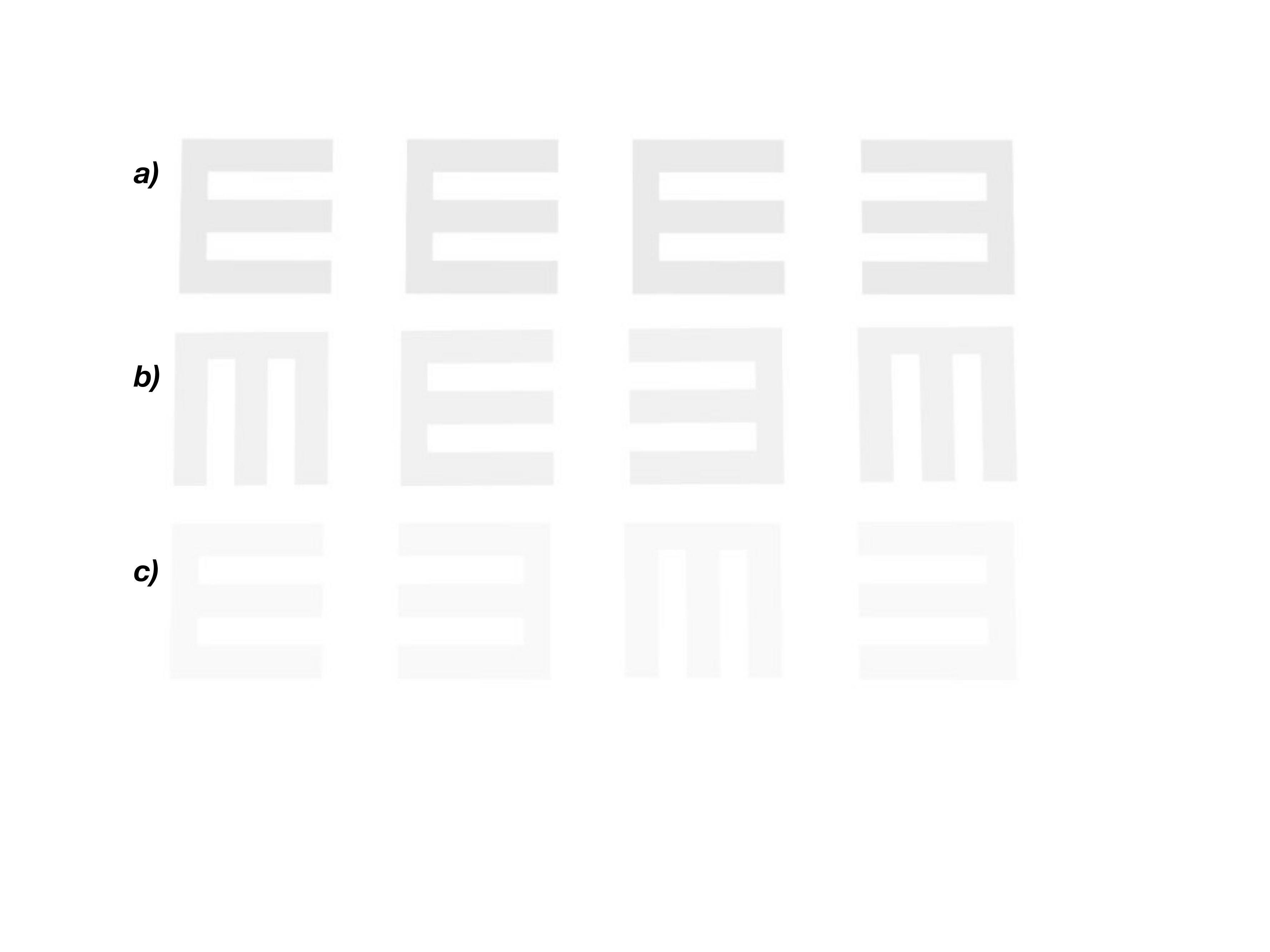}
 \caption{View of the contrast test, \textit{a)} shapes rendered with an $\alpha$ value of $30$, \textit{b)} $20$, and \textit{c)} $10$.}
 \label{fig:ContrastScreenshot}
\end{figure}

\paragraph{Perceived contrast (\textit{image quality}):} The ability to differentiate between various levels of gray is particularly applicable to the medical setting where images such as slices from MRI and CT scans are visualized in grayscales.
Similar to the previous metric, the users are positioned at a fixed distance away from the virtual test material.
The test contains four shapes with distinctive rotations (up, down, right, and left) and the users are tasked with reducing the visibility of the shapes until they are not able to discern their rotation. Increasing transparency is achieved by reducing the alpha channel value of the shapes materials as shown in Fig. \ref{fig:ContrastScreenshot} for three different alpha channel values: \textit{\textbf{a)}} at $15$, \textit{\textbf{b)}} at $10$ and \textit{\textbf{c)}} at $6$.

\paragraph{Head mounting ease (\textit{secondary features}):} This feature is primarily influenced by the amount of adjustable parts. The complexity of head mounting affects the time it takes each wearer to mount the HMD, hence it is of great relevance when multiple people need to wear the device successively. To asses this property, the users receive verbal instructions on the proper way to wear and adjust the HMD, wearing time is measured from the moment they hold the HMD to the time they report the completion of the step.
    
\paragraph{Compatibility with glasses (\textit{comfort}):} Approximately $64\%$ of the United States population wear eyeglasses~\cite{visioncouncil}. Therefore, the compatibility of any VR HMD that is considered for large scale deployment with optical glasses is necessary. In a questionnaire, users are asked to report the comfort of the HMD with their glasses.

\section{Experimental results and evaluation}
The experiment is designed so that we exclude certain HMDs from the tests based on the results from the previous category or test. This allows us to consider a broader mass of HMDs at the beginning and reducing the number for the more detailed tests.

The evaluated HMDs include: \textit{HTC Vive (HTC Corporation, Taoyuan, Taiwan), HTC Vive Pro (HTC Corporation, Taoyuan, Taiwan), Oculus Rift S (Facebook, Menlo Park, CA, US), Samsung Odyssey, Samsung Odyssey+ (Samsung, Seoul, South Korea), HP WMR headset (HP Inc., Palo Alto, CA, US), Acer WMR headset (Acer Inc. New Taipei, Taiwan), Lenovo Explorer (Lenovo Group Limited, Beijing, China), Dell Visor (Dell, Round Rock, TX, US), and Pimax 5k (Pimax, Shanghai, China)}. In the following evaluation we strictly rely on the metrics introduced in Sec.~\ref{sec:method} to evaluate the HMDs. Other information that are available about the HMDs are, if at all, only used as additional discussion to our results. 

\begin{table*}[t]
  \caption{Overview of metrics that were taken from the data sheet and measurements}
  \label{tab:big_table}
	\centering%
  \begin{tabu} to \textwidth {%
	ll%
	*{9}{l}%
	}
  \toprule
   \textbf{HMD} & Resolution &   Field of View & IPD & Integrated & Connectivity \\
   & (per eye) &&& audio\\
  \midrule
    Samsung Odyssey+ & $1440\times1600$ & $110^\circ$ & Yes & Yes & HDMI\\
	HTC Vive Pro & $1440\times1600$ & $110^\circ$  & Yes & Yes & Display Port, 3x power outlet \\
	Oculus Rift S & $1280\times1440$ & $>110^\circ$\textsuperscript{3} & Digital & Yes\textsuperscript{2} & Displayport or mini displayport\\
	Pimax 5k & $\mathbf{2560\times1440}$ & $\mathbf{170^\circ}$ & Yes & No & Displayport, 3x power outlet\\
	HTC Vive & $1080\times1200$ & $110^\circ$ & Yes & Yes\textsuperscript{1} & HDMI, 3x power outlet\\
	Samsung Odyssey & $1440\times1600$ & $110^\circ$ & Yes & Yes & HDMI, Bluetooth\\
	Lenovo Explorer & $1440\times1440$ & $110^\circ$ & No & No & HDMI, Bluetooth\\
	Dell Visor & $1440\times1440$ & $110^\circ$ &  No & No & HDMI, Bluetooth\\
	Acer WMR & $1440\times1440$ & $100^\circ$ & No & No & HDMI, Bluetooth\\
	HP WMR & $1440\times1440$ & $95^\circ$ & No & No & HDMI, Bluetooth\\
	& & & & & \\
	& & & & & \\
	& & & & & \\
	 & Heat & Torque & Weight & Weight to & Hygiene & Brightness & Color \\
	& measurement & on neck (Nm) & (g) &torque ratio  & &($\frac{cd}{m^2}$) & accuracy\\
	& ($^\circ$C)&&&($\frac{g}{Nm}$) & & & ($\Delta\, e$)\\
  \midrule
    Samsung Odyssey+ & $31.4$ & $0.708$ & $610$ & $861.6$ & Replacable & $127.5$ & $11.8$ \\
    HTC Vive Pro & $37.4$ & $0.777$ & $814$ & $\mathbf{1047.6}$ & Replacable & $133.3$ & $\mathbf{6.5}$ \\
    Oculus Rift S & $32.3$ & $0.737$ & $584$ & $792.4$ & - & $80.5$ & $11.7$ \\
    Pimax 5k & $28.6$ & $0.902$ & $593$ & $657.4$ & Replacable & $44.4$ & $8.1$  \\
    HTC Vive & $\mathbf{27.6}$ & $0.866$\textsuperscript{1} & $732$\textsuperscript{1} & $845.3$ & Replacable & $\mathbf{190.5}$ & $8.9$ \\
    Samsung Odyssey & $30.8$ & $0.837$ & $662$ & $790.8$ & Wipeable, replacable & $126.5$ & $7.1$ \\
    Lenovo Explorer & $37.6$ & $0.502$ & $\mathbf{419}$ & $834.6$ & Replacable & $88.7$ & $15.5$ \\
    Dell Visor & $32.2$ & $0.635$ & $613$ & $965.4$ & Replacable & $76.2$ & $7.7$ \\
    Acer WMR & $35.2$ & $\mathbf{0.494}$ & $464$ & $939.3$ & Replacable & $76.6$ & $28.6$ \\
    HP WMR & $33.8$ & $0.647$ & $530$ & $819.2$ & Replacable & $101$ & $16.6$\\

  \midrule
  \multicolumn{2}{l}{1 With the Deluxe Audio Strap}\\
  \multicolumn{5}{l}{2 The speakers are open, the people in the environment also hear the audio}\\
  \multicolumn{8}{l}{3 Diagonal FoV, horizontal field of view was measured to be $94^\circ$: \textit{http://doc-ok.org/?p=1414}}\\
  \bottomrule
  \end{tabu}%
\end{table*}

\subsection{Data sheet} \label{subsec:datasheetResults}
The measurements associated to the metrics of this category are presented in Table~\ref{tab:big_table}. \textit{Oculus Rift S} has a digital \textit{IPD} adjuster as it uses only one display instead of two separate displays. While the physical \textit{IPD} adjuster moves the lenses closer to each other or farther apart, the digital one only translates the two images on the plane of display towards the edges or closer to each other. Since the distance of the lenses remain constant, digital \textit{IPD} adjustment results in a diminished effect. It is important to note that since all HMDs require a USB 3.0 port, therefore it is excluded from the list under the \textit{Connectivity} column.

\subsection{Measurements}

\subsubsection{Neck strain \& total weight}
We measured the \textit{torque} around the neck for all HMDs. The \textit{torque} measurements are presented in Table~\ref{tab:big_table}. The fifth column in the lower half of the table represents gram per Newton-meter, \textit{i. e.} the ratio between the \textit{weight} and \textit{torque}. A higher value refers to a lower \textit{torque to weight} ratio. While the absolute \textit{torque} applied around the neck is an important measure related to the user comfort, the \textit{weight to torque ratio} allows us to compare the ergonomic design across different HMDs.

\subsubsection{Brightness \& Color accuracy}
The result of the display calibration are shown in Table~\ref{tab:big_table}. For \textit{brightness}, a higher value is more desirable, while for the color deviation a lower $\Delta\, e$ is associates with more accurate color representation, hence is more desirable.

\subsubsection{Heat management} 

\begin{figure}[tb]
 \centering 
 \includegraphics[width=\columnwidth]{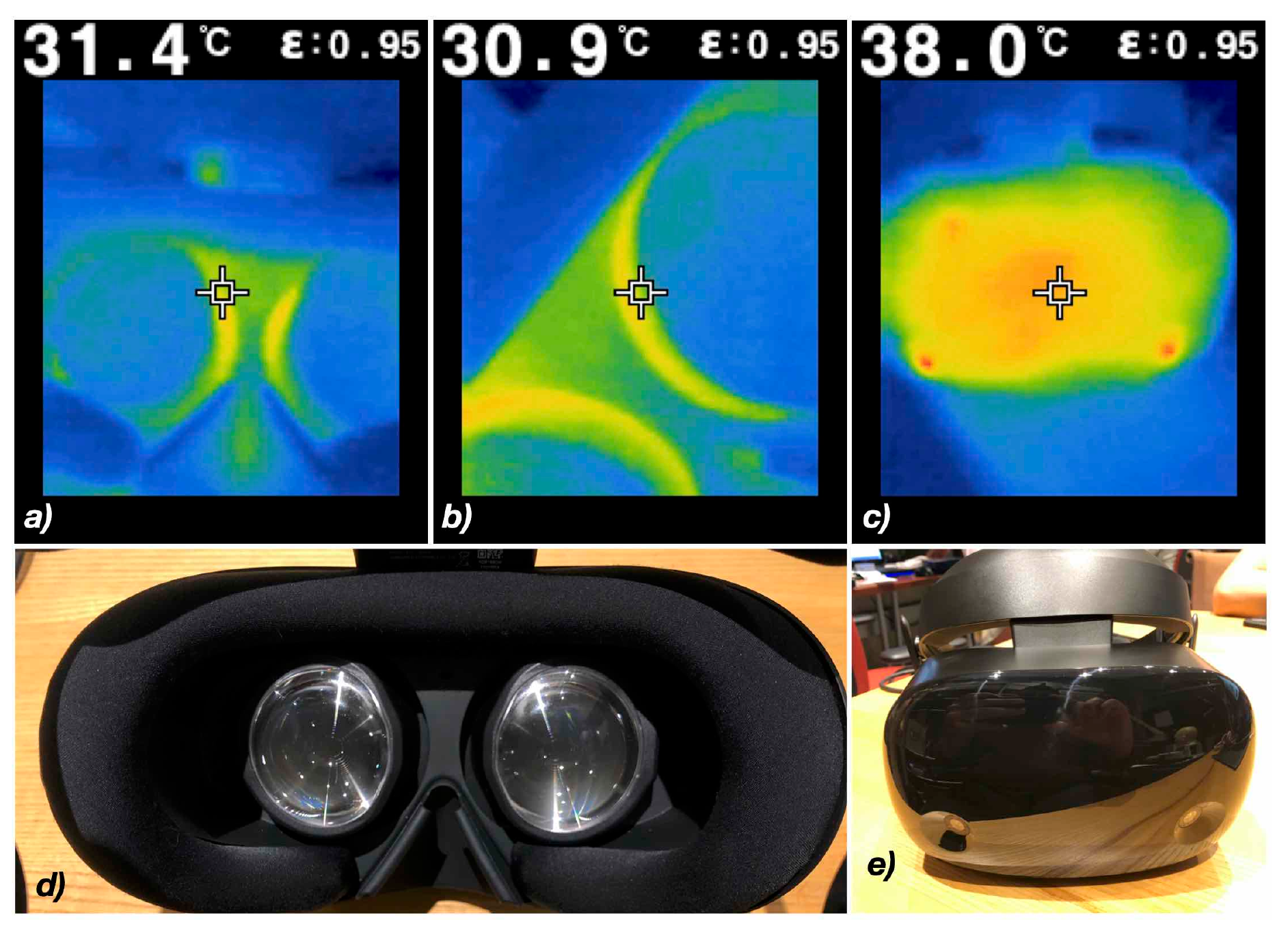}
 \caption{\textit{Samsung Odyssey+}, \textbf{(a, b)} showing the hottest spot on the lens-side of the HMD, and \textbf{(c)} the hottest spot on the front side of the HMD. The corresponding views of the heat images are shown in \textbf{(d, e)}.}
 \label{fig:heatOdy}
\end{figure}

\begin{figure}[tb]
 \centering 
 \includegraphics[width=0.96\columnwidth]{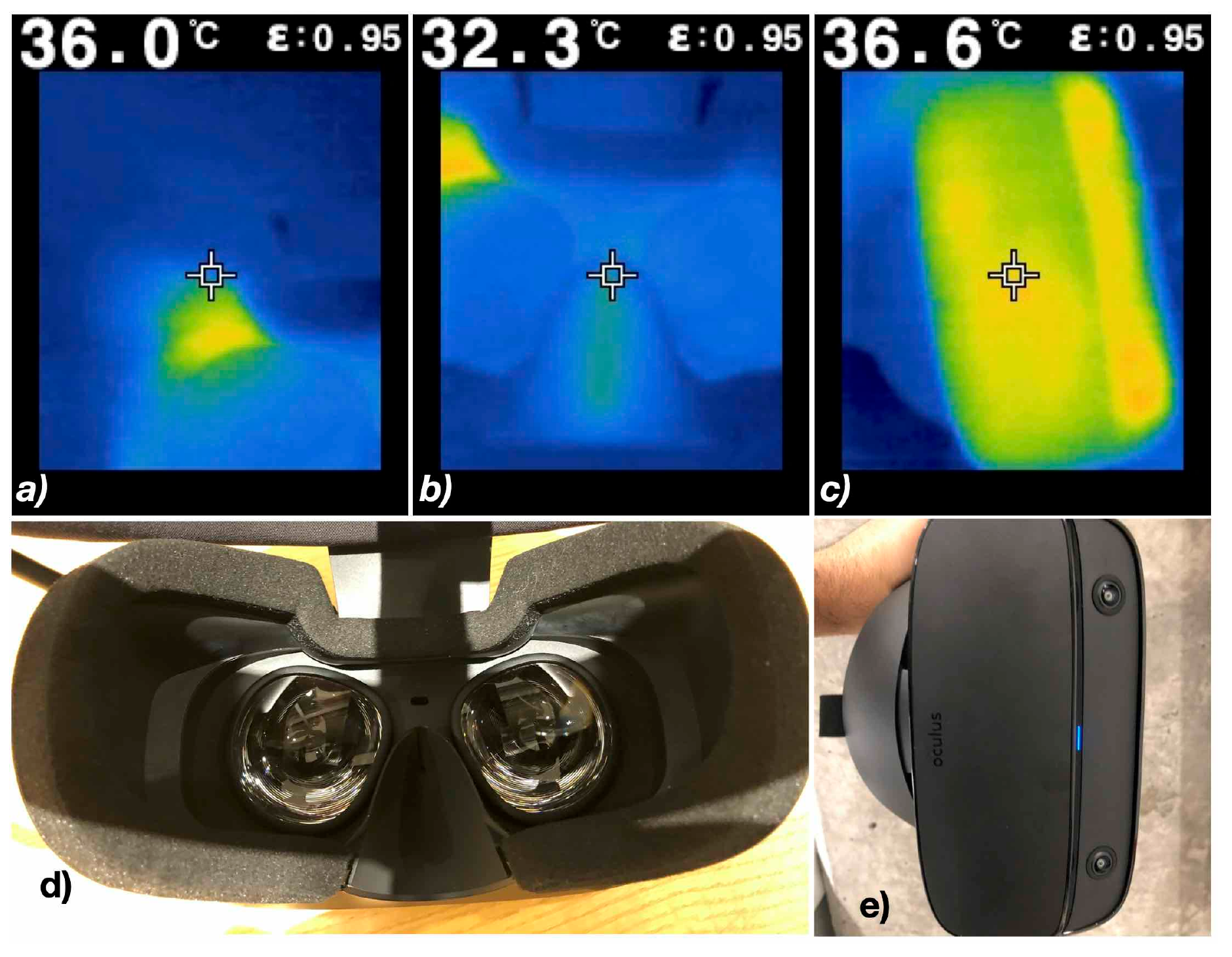}
 \caption{\textit{Oculus Rift S}, \textbf{(a)} showing the hottest spot on the 'face-side' of the HMD, \textbf{(b)} the heat image from between the lenses, and \textbf{(c)} the hottest spot on the front side of the HMD. The corresponding views are shown in \textbf{(d, e)}.}
 \label{fig:heatRift}
\end{figure}

\begin{figure}[tb]
 \centering
 \includegraphics[width=\columnwidth]{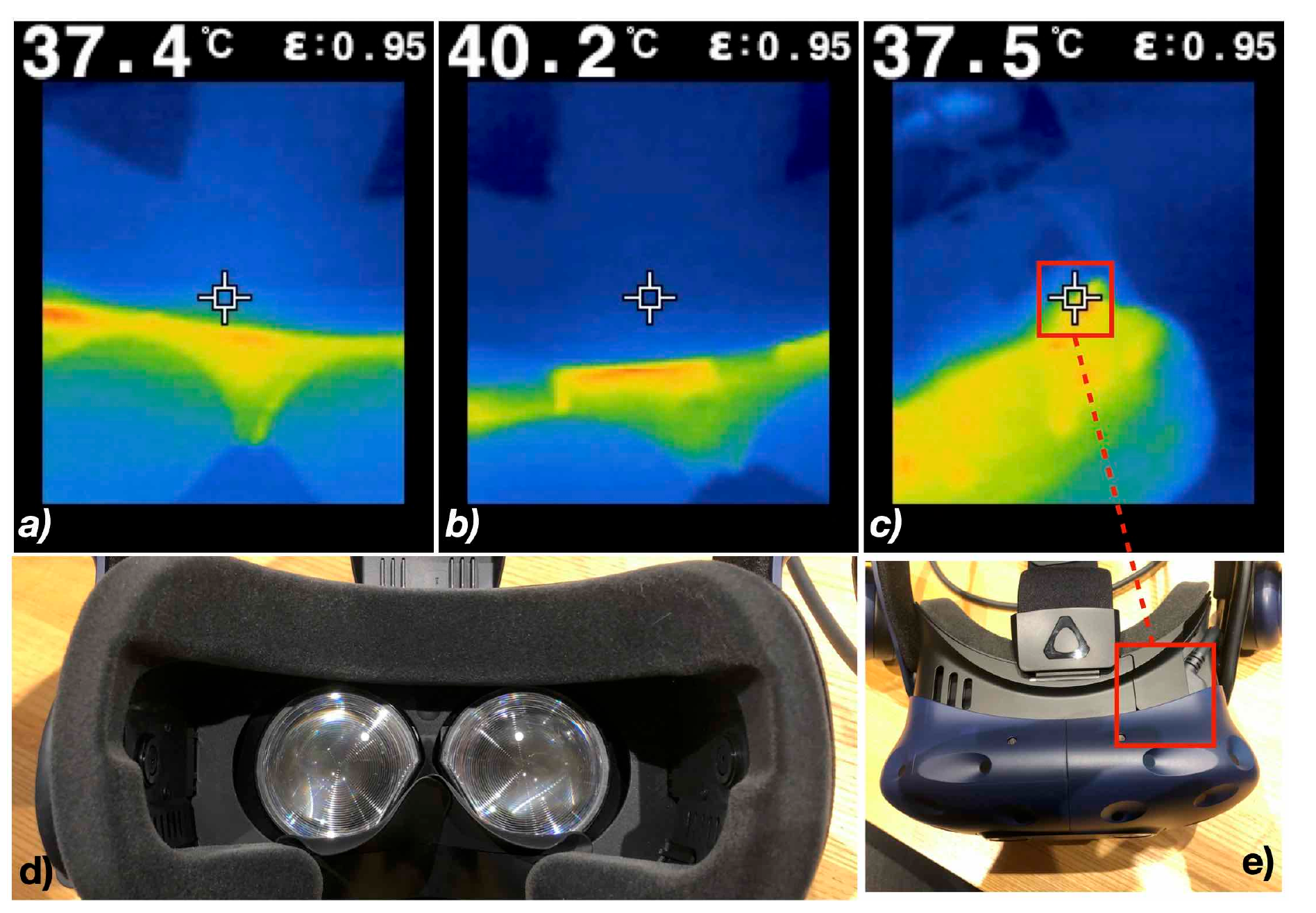}
 \caption{\textit{HTC Vive Pro}, \textbf{(a)} showing the hottest spot in the center and above the lenses, \textit{b)} a spot with higher temperature that is acquired from a small cavity on the top side, therefore excluded from the measurements, and \textbf{(c)} the hottest spot on the front side of the HMD. The corresponding views are shown in \textbf{(d, e)}.}
 \label{fig:heatVive}
\end{figure}

In Fig.~\ref{fig:heatOdy},~\ref{fig:heatRift}, and~\ref{fig:heatVive} we show the measurements from the hottest spots inside and outside the VR devices. For consistent evaluation among all devices, we measured the hottest spot at the center of the HMD, and listed all measurements in Table~\ref{tab:big_table}. We also encountered spots with higher temperature away from the center of the device, \textit{e.g.} from the \textit{HTC Vive Pro} we recorded a temperature of $40.2$\textdegree located at the top of the HMD, and from the top left of the \textit{Oculus Rift S} we recorded the temperature $36.0$\textdegree.


\subsubsection{Hygiene}
\textit{Samsung Odyssey} is the only HMD that is identified as wipeable (Table~\ref{tab:big_table}). All HMDs except \textit{Oculus Rift S} have replaceable facial interfaces.

\subsection{Multi-user study}
In the multi user study, $27$ users were invited to evaluate \textit{readability} and \textit{contrast perception} inside a VR application, and were asked to fill out a questionnaire. The user experiments were only conducted on the three best performer HMDs given the criteria evaluated based on their datasheet and lab measurements. These devices include \textit{Samsung Odyssey+}, \textit{HTC Vive Pro}, and \textit{Oculus Rift S}. To counteract the learning bias, the order in which the users wore the HMDs was randomized.

The space dedicated to the VR environment was $2 \times 3$\,m$^2$ in size, users could move freely within this area. We turned off the \textit{HTC Lightouse} boxes when they were not used to prevent interference with the tracking of the other HMDs. The VR environment ran on an \textit{Alienware (Dell, Round Rock, TX, US)} laptop with an \textit{Intel i7-7700HQ\,k} cpu, \textit{NVIDIA GTX 1070} graphics card, and a $16$\,GB RAM.

\subsubsection{Text readability}
\begin{figure}[tb]
 \centering
 \includegraphics[width=\columnwidth]{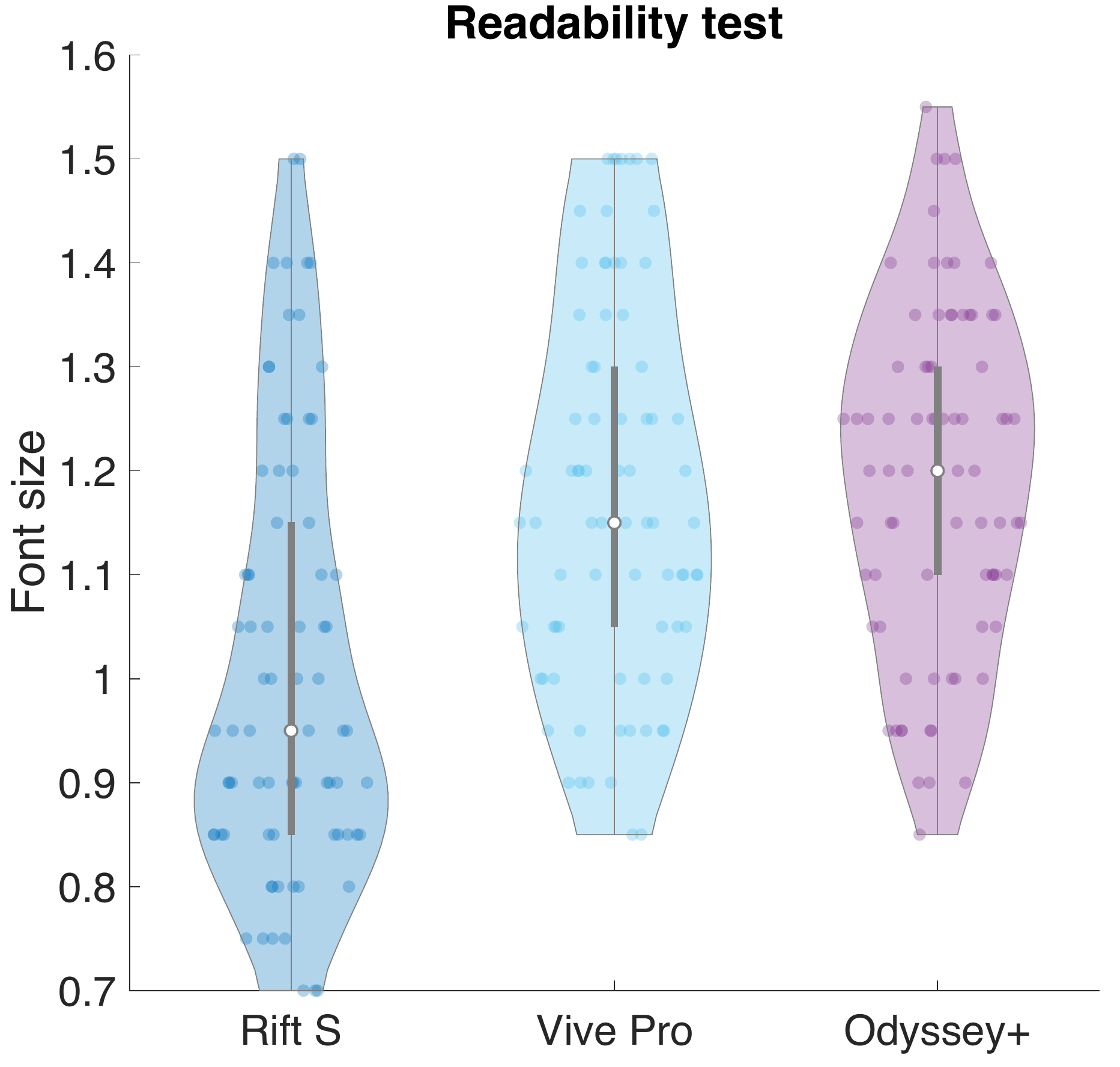}
 \caption{Violinplot showing the results of the readability test, where lower font size is related to better readability. \textit{Oculus Rift S} outperformed the other devices in text readability. The other two HMDs performed similarly, with the \textit{HTC Vive Pro} demonstrating readability at slightly lower font sizes.}
 \label{fig:readability}
\end{figure}

The font sizes reported by users are shown in the plot in Fig.~\ref{fig:readability}, with a more detailed overview in Table \ref{tab:readability}. \textit{Oculus Rift S} demonstrated a lower average, median and minimum font size compared to the other two HMDs.

\begin{table}[t]
  \caption{Readability test outcome, where values indicate the smallest font size that was readable by users}
  \label{tab:readability}
	\centering%
  \begin{tabu} to \textwidth {%
	llllll%
}
  \toprule
   \textbf{HMD} & Avg & Stdev & Min & Max & Med\\
  \midrule
    Samsung Odyssey+ & $1.19$ & $\mathbf{0.16}$ & $0.85$ & $1.5$ & $1.2$\\
    Oculus Rift S & $\mathbf{1.01}$ & $0.21$ & $\mathbf{0.7}$ & $1.5$ & $\mathbf{0.95}$\\
    HTC Vive Pro & $1.17$ & $0.19$ & $0.85$ & $1.5$ & $1.15$\\
  \bottomrule
  \end{tabu}%
\end{table}

\subsubsection{Contrast} \label{subsubsec:contrast}

\textit{Samsung Odyssey+} exhibited an average contrast of $3.385$ with a standard deviation of $0.913$. This is the lowest perceivable contrast among the other HMDs. \textit{HTC Vive Pro} ranked second with an average contrast of $5.115$ and a standard deviation of $0.526$. Finally, the \textit{Oculus Rift S} yielded an average contrast of $6.038$ with the standard deviation of $0.445$.

\subsubsection{HMD wearing time}
Each participant received verbal introduction for wearing each HMD. These instructions included every component that needed adjustment, optimal position of the device on the face, and the correct usage of the controllers. Subsequently, the study proctor measured the required time for each participant to wear the HMD. 

\begin{table}[t]
  \caption{The time it took the users to mount each HMD in seconds}
  \label{tab:timeHMD}
	\centering
  \begin{tabu} to \textwidth {
	llllll
}
  \toprule
   \textbf{HMD} & Avg & Stdev & Min & Max & Med\\
  \midrule
    Samsung Odyssey+ & $\mathbf{50.85}$ & $\mathbf{23.69}$ & $\mathbf{16}$ & $\mathbf{126}$ & $\mathbf{52}$\\
    Oculus Rift S & 114.58 & 75.04 & 25 & 356 & 95.5\\
    HTC Vive Pro & 112.27 & 68.29 &  53 & 297 & 91\\
  \bottomrule
  \end{tabu}
\end{table}

\begin{figure}[tb]
 \centering
 \includegraphics[width=\columnwidth]{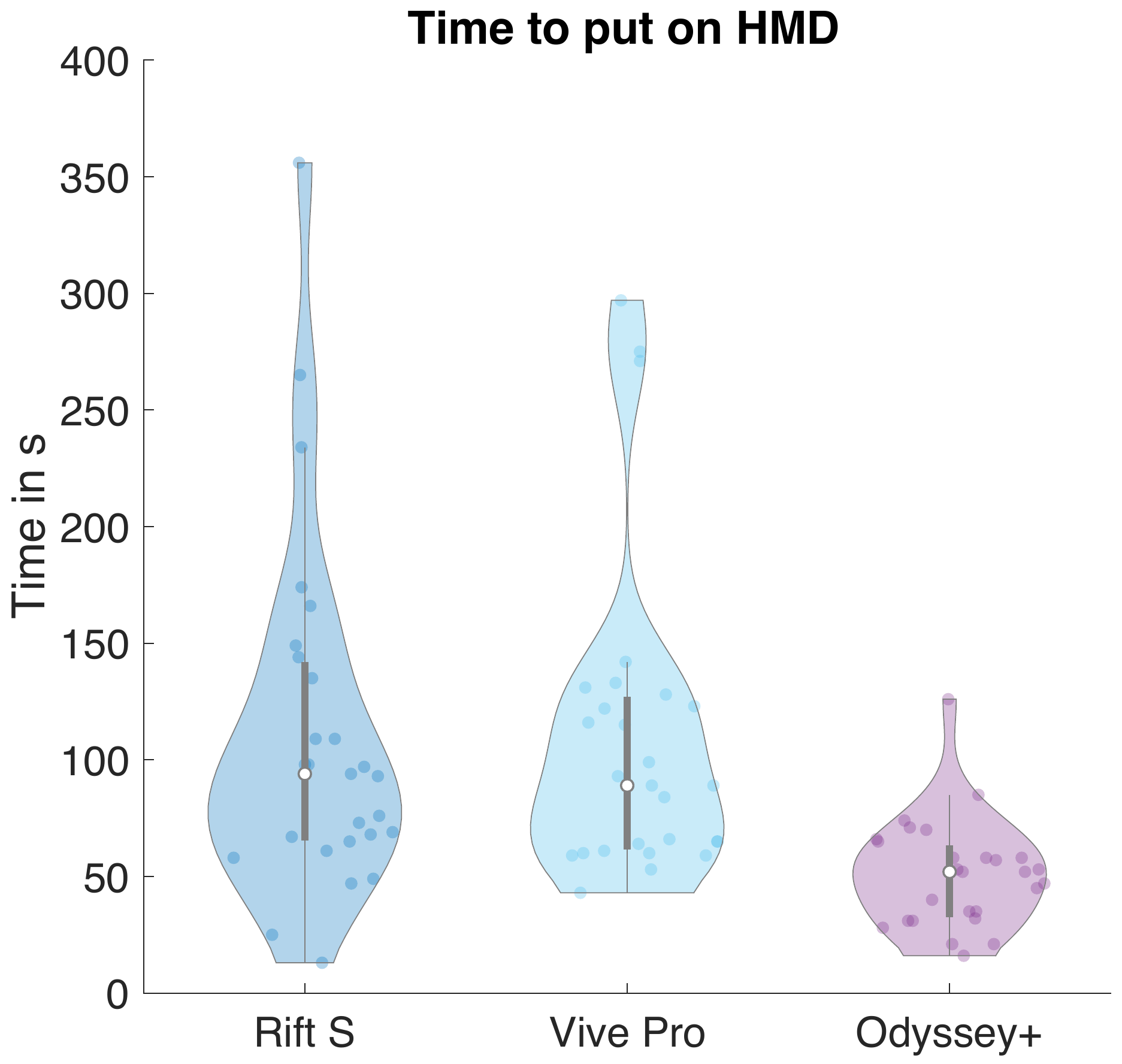}
 \caption{Violinplots showing the measured time to wear each HMD}
 \label{fig:time}
\end{figure}

The results are shown in Fig.~\ref{fig:time}. The \textit{Samsung Odyssey+} took the shortest time to wear. The other two HMDs exhibited very similar outcomes, with the \textit{HTC Vive Pro} being slightly faster.

\subsubsection{Compatibility with glasses}
Out of the $27$ users, $11$ wore glasses. The results for these participants are reported in Table~\ref{tab:glassesComfort}. \textit{HTC Vive Pro} was characterized as the most compatible HMD with glasses. \textit{Oculus Rift S} differed by only one user which perceived it as incompatible and uncomfortable. The \textit{Samsung Odyssey+} was perceived considerably more uncomfortable.

\begin{table}[ht]
  \caption{Users wearing glasses report how comfortable they experienced each HMD}
  \label{tab:glassesComfort}
  	\centering
  \begin{tabu} to \textwidth {
	llllll
}
  \toprule
   \textbf{HMD} & Comfortable & Okay & Uncomfortable\\
  \midrule
    HTC Vive Pro & $\mathbf{9}$ & $1$ & $\mathbf{1}$\\
    Oculus Rift S & $8$ & $1$ & $2$\\
    Samsung Odyssey+ & $4$ & $3$ & $4$\\
  \bottomrule
  \end{tabu}
\end{table}

\subsubsection{Comfort}

\begin{figure*}[t]
    \centering
    \subfigure[]{\includegraphics[scale=.3]{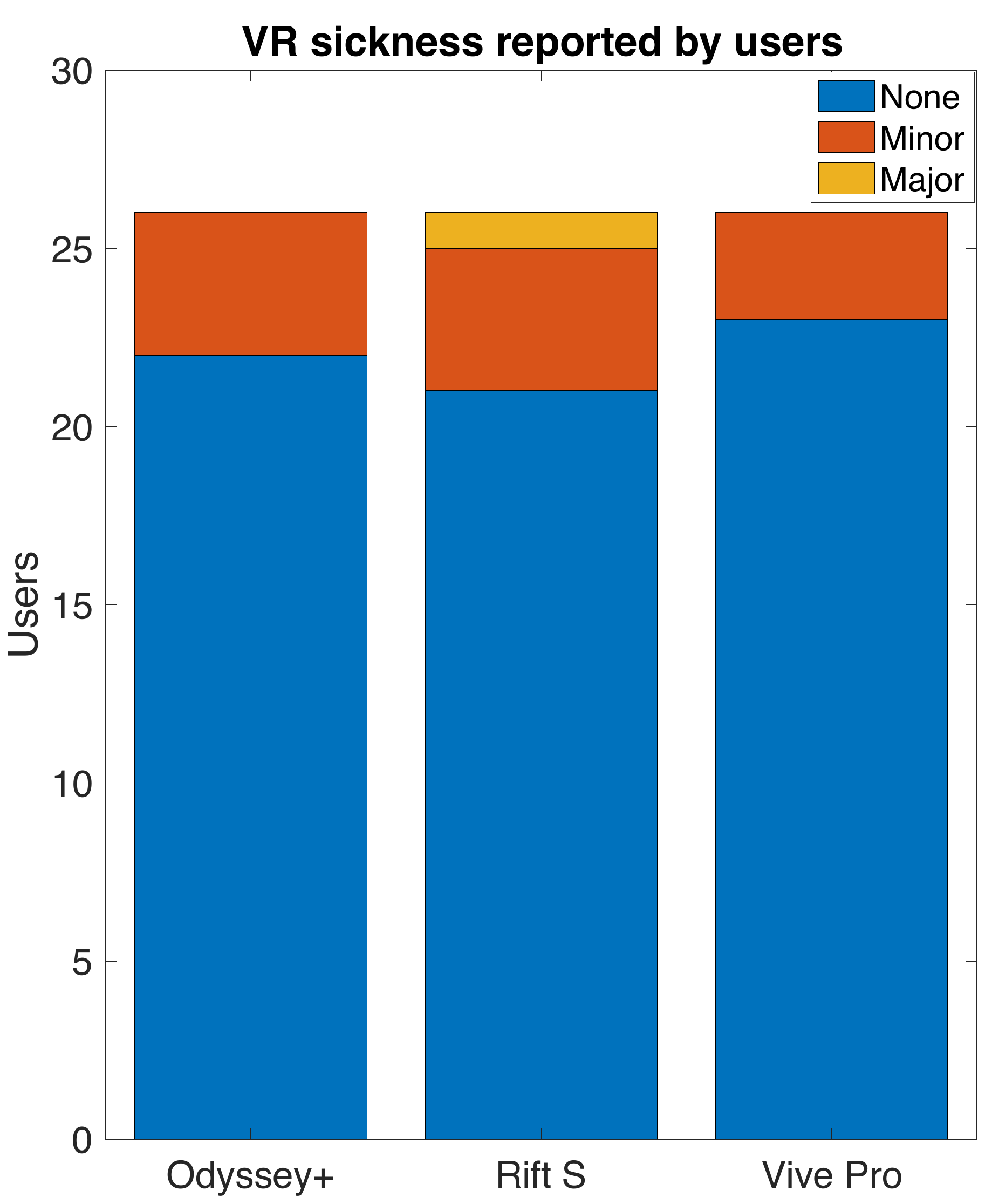}}
    \subfigure[]{\includegraphics[scale=.3]{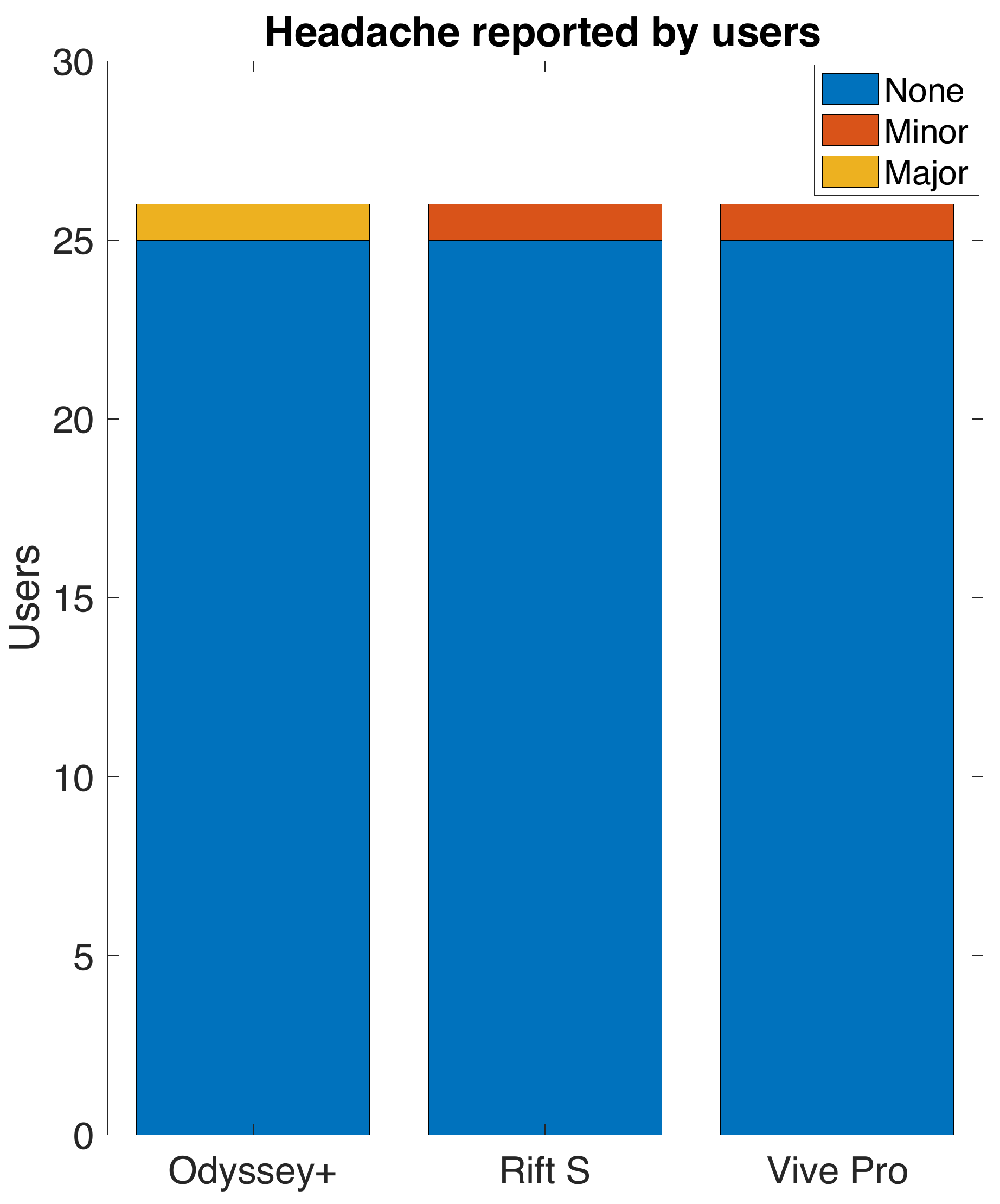}}
    \subfigure[]{\includegraphics[scale=.3]{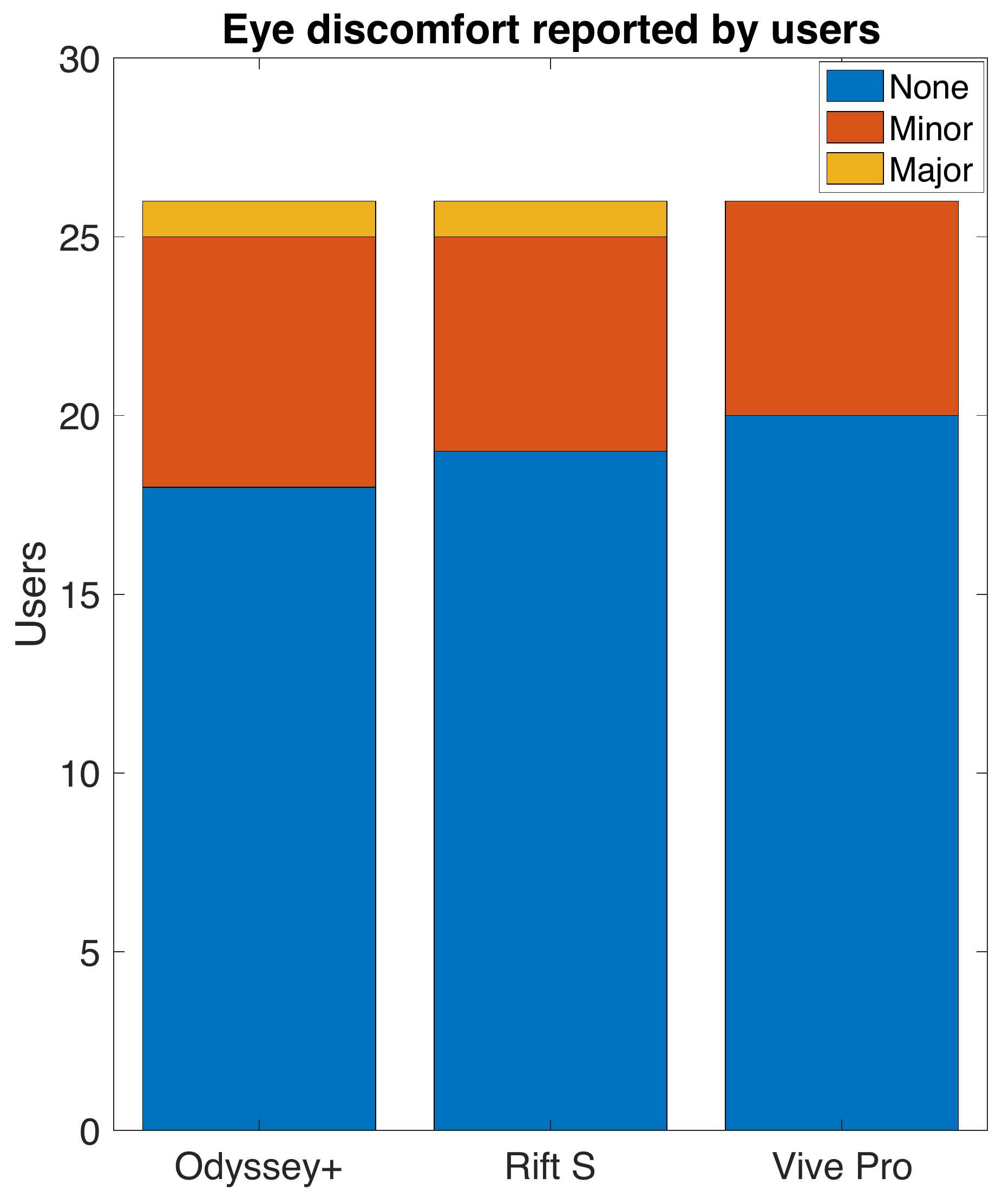}}
    \caption{Charts (a), (b), and (c) illustrate the distribution of users who reported VR sickness, headache, and eye discomfort respectively; \textbf{blue} denotes a user who reported no complaints, \textbf{orange} for a user with mild or minor discomfort, and \textbf{yellow} users who experienced major discomfort.}
    
    
    \label{fig:comfortFigs}
\end{figure*}

Fig.~\ref{fig:comfortFigs} are results of the user-questionnaires. Within the questionnaires, we evaluated the properties relevant to user comfort and satisfaction for each HMD.
To gather information about comfort, participants responded whether they experienced \textit{VR sickness}, \textit{headache}, or \textit{eye discomfort}. The data shows only minor discrepancy across different HMDs. Among the three VR devices, \textit{HTC Vive Pro} has the least number of participants reporting any sort of the aforementioned discomforts. 

\section{Discussion}
As aforementioned we choose the top three HMDs based on their datasheet and lab-measurements. The last four devices as shown in Table~\ref{tab:big_table} are excluded solely due to their lack of an adjustable \textit{IPD}. Furthermore our results show that \textit{Samsung Odyssey+} and \textit{HTC Vive Pro} are overall improvements on the \textit{Samsung Odyssey} and \textit{HTC Vive}, which is why we only consider the former devices for the multi-user study. The \textit{Pimax 5k} was excluded as a direct result of the extreme \textit{neck-strain} it exerted, in addition it had the lowest $\frac{g}{Nm}$ ratio, from all measured HMDs. This is reasonable when looking at the head strap of the \textit{Pimax 5k}, where most HMDs have sophisticated ergonomic head mounts that distribute the \textit{weight} over the whole head, it consists of simple elastic bands.
For an in-depth analysis, we mostly limit the discussion in this section to the top three HMD devices that were used during the multi-user study.

\subsection{Data sheet}
The upper half of Table~\ref{tab:big_table} highlights the HMDs that performed best given the following criteria. 

The \textit{Samsung Odyssey+} has a horizontal\textit{FoV}of $110^{\circ}$, uses two AMOLED displays with a \textit{resolution} of $1440\times1600$ pixels per eye, and connects to the computer via an HDMI and USB cable. The \textit{HTC Vive Pro} has the same FoV, uses two OLED displays with the same \textit{resolution} per eye, however it connects to the computer through display-port and USB cable and requires three power sockets. The \textit{Oculus Rift S} has a diagonal\textit{FoV}of $110$\textdegree, uses one LCD display with a lower \textit{resolution} of $1280\times1440$ pixels per eye, and connects to the computer via a display-port and USB cable. The\textit{FoV} specification given by \textit{Oculus} only states a diagonal \textit{FoV} that is supposedly slightly larger than $110$\textdegree. 
No standardized procedure to measure \textit{FoV} is established, the \textit{FoV} the user conceives substantially depends on the distance and correct alignment between eye and lens and is therefore influenced by 
the shape of the users face and even the correct \textit{IPD} adjustment. These factors aggravate the direct comparison of \textit{FoV} stated by the manufacturer.

Similar to the \textit{Samsung Odyssey+}, the \textit{Oculus Rift S} uses a Simultaneous Localization and Mapping (SLAM)-based tracking, therefore estimates its pose in 3D-space without relying on an external tracking system, the \textit{Samsung Odyssey+} however employs two RGB cameras whereas the \textit{Oculus Rift S} utilizes five RGB cameras thus laying a foundation for a more stable tracking system.



Relying merely on Bluetooth as an external factor is undesirable, since not all computers have Bluetooth capabilities. Furthermore, incompatible Bluetooth adapters can cause interruptions and affect the performance. Both HMDs from \textit{HTC} as well as the \textit{Pimax 5k} require three additional power connections, where one is to power the HMD itself, and the other two are for the \textit{HTC Lighthouse} tracking system that consists of two boxes that are placed at the two corners of the VR space. While this is generally considered as superior in terms of tracking stability, the external setup drastically impacts the usability of the system, especially when compared to the HMDs which use an inside-out marker-less SLAM strategy for tracking.

\subsection{Measurements}

The \textit{HTC Vive Pro} is the heaviest device and has the best \textit{weight to torque ratio}, which advocates a good ergonomic headband design; notwithstanding, from all three devices, it exerts the highest \textit{force on the neck}.

The \textit{Samsung Odyssey+}, ranked second in \textit{weight} and \textit{weight to torque ratio} however measured the best results in \textit{neck-strain}.
\textit{Oculus Rift S} has the lowest \textit{weight} but ranks second in terms of \textit{neck-strain} and has consequently the lowest \textit{weight to torque ratio}.


The \textit{HTC Vive Pro} has the highest \textit{brightness} with the lowest $\Delta\,e$, while the \textit{Samsung Odyssey+} and the \textit{Oculus Rift S} have a comparable \textit{color accuracy}, the latter has a lower \textit{brightness}.

The \textit{HTC Vive Pro} has the brightest display with $133.3$\,nits and most accurate color representation, $\Delta\,e$ of $6.5$. \textit{Samsung Odyssey+} has a slightly lower \textit{brightness} measurement of $127.5$\,nits and a substantially higher - and therefore worse - $\Delta\,e$ of $11.8$, whereas \textit{Oculus Rift S} has a similar $\Delta\,e$ of $11.7$, but a considerably lower \textit{brightness} of $80.5nits$. In regards to \textit{brightness} and \textit{color accuracy}, the \textit{HTC Vive Pro} performs best, indicating superior display quality.

Our results indicate the \textit{Samsung Odyssey+} has the least \textit{heat} development inside the HMD, with the hottest measurement being $31.4^{\circ}$. The \textit{Oculus Rift S} has a marginally higher measurement of $32.3^{\circ}$ and the \textit{HTC Vive Pro} has highest measurement with $37.4^{\circ}$. As mentioned above the latter two HMDs had hotter areas, facilitating our decision against using them as they were either on the edge of the facial interface or underneath another layer of cover. The differences between the first two HMDs are negligible, only the \textit{HTC Vive Pro} has a substantially higher measurement.

Since none of the final three HMDs we tested had a \textit{wipeable} facial interface, it is important that the interface can be switched out for a new one. Likewise, covers for the facial interface that are new or user-specific can be used as well. \textit{HTC Vive Pro} and \textit{Samsung Odyssey+} have \textit{replaceable} interfaces, while the \textit{Oculus Rift S} has a non-replaceable facial interface. 

\subsection{Multi-user study}
Considering that the \textit{Oculus Rift S} has the lowest \textit{resolution} and performed worst in the \textit{brightness} and \textit{color accuracy} test it is very surprising that it scored the distinctly best \textit{readability} result in this test. It has the lowest average, lowest minimal and median value, as shown in table \ref{tab:readability}, while the distribution can be seen in figure \ref{fig:readability}. The other two HMDs have a similar performance, with the \textit{HTC Vive Pro} performing slightly better. 

As described in section \ref{subsubsec:contrast} the \textit{Samsung Odyssey+} had the best perceivable \textit{contrast}, allowing users to discern shapes at an alpha value of $3.385$, whereas the \textit{HTC Vive Pro} has an average value of $5.115$ and the \textit{Oculus Rift S} $6.038$. 
The results for the latter device might be justified by the LCD screen, which are known to have an inferior \textit{contrast} to OLED displays.

HMD \textit{mounting-time} reflects on the general usability of the HMD and convenience for the wearer. It was hypothesized that the \textit{Samsung Odyssey+} performs best in this category, as it has the least adjustable components when mounting the HMD. Table \ref{tab:timeHMD} shows that on average, compared to the other two HMDs, users were twice as fast when mounting and adjusting the device.
The \textit{HTC Vive Pro} is the bulkiest from all HMDs and requires many adjustments before using, so it was to be expected that users took longer to mount the device. The \textit{Oculus Rift S}, on the other hand, requires a substantial amount of time to be put on as users have to digitally adjust the \textit{IPD} in the software menu. In this study, prior to use of the \textit{Oculus Rift S}, the proctor handed the HMD to the users with the \textit{IPD} menu already opened. It has to be taken into account that new users need more time to adjust the HMD than seasoned users of VR, which explains the high standard deviation. Nonetheless, we believe that the trend that manifests itself in our data will remain as users gain experience with the devices, especially considering the interaction with the software in the \textit{Oculus Rift S}. 

Users found the \textit{HTC Vive Pro} as the most compatible HMD with glasses, followed by the \textit{Oculus Rift S} and \textit{Samsung Odyssey+}.
This outcome was not surprising  as the former two devices allow users to adjust the distance between the lenses and their eyes, thereby increasing space for glasses. 

The results from the questionnaires in Fig.~\ref{fig:comfortFigs} show that the majority of the users did not experience \textit{VR sickness}, \textit{headache}, or \textit{eye discomfort}. Nonetheless, a slight trend is discernible and is henceforth discussed. Regarding \textit{VR sickness}, for the \textit{HTC Vive Pro}, $23$ users reported none, $3$ minor, and $0$ users major \textit{VR sickness}. For the \textit{Samsung Odyssey+}, $22$ users reported none, $4$ minor, and $0$ major \textit{VR sickness}. Lastly, for the \textit{Oculus Rift S}, $21$ reported none, $4$ minor, and $1$ major \textit{VR sickness}. The discernible trend shows that users found the \textit{HTC Vive Pro} slightly more comfortable than the other two HMDs. The same trend is visible in Fig.~\ref{fig:comfortFigs} \textit{(b) and (c)}, where users report more \textit{headache} and \textit{eye discomfort} on the other two HMDs, with the \textit{Samsung Odyssey+} performing worse than the \textit{Oculus Rift S}. We hypothesize that for the latter HMD, the digital \textit{IPD} adjustment influenced the comfort for a subset of users, as the misaligned lenses cause severe discomfort for people whose \textit{IPD} are far outside the average. 

The additional users' remarks align with the results of the multi-user study and provide further insight into the discomfort that users experienced with the \textit{Samsung Odyssey+}. The most prominent note from users' feedback is the pressure on the face and head they experienced while wearing \textit{Samsung Odyssey+}, leading to major dissatisfaction and slight pain. Furthermore, users reported that the edges inside the \textit{Samsung Odyssey+} have a noticeable aliasing effect, especially when tilting the head. Another visual artifact was noted that occurred during fast movement of the head, where the shadow of the objects shown on the screen lagged behind. Lastly, a flickering of the letters during the readability test was reported. The tracking of the \textit{Samsung Odyssey+} was reported as the least stable, followed by the \textit{Oculus Rift S} and then the \textit{HTC Vive Pro}. The instability was particularly noticeable when the users translated their heads to the side. When examining the HMDs ourselves, we were able to confirm most of the aforementioned reports.

\section{Conclusion}
In this work, we presented metrics that are relevant for the evaluation and assessment of VR HMDs for use in different applications. We categorized, measured, and systematically evaluated various VR designs and technologies. Each HMD manifested its own strengths and weaknesses. Despite the average performance in the readability and contrast perception tests, we found the \textit{HTC Vive Pro} to overall excel, especially in regard to ergonomics and comfort. We observed that several metrics that were measured in this work had direct influence on user comfort, these include: \textit{image quality}, \textit{heat} development, tracking stability, \textit{weight}, and compatibility with glasses.




\bibliographystyle{abbrv-doi}

\bibliography{template}
\end{document}